\documentclass[review,12pt]{elsarticle}
\usepackage{geometry}
\usepackage{amsmath}
\usepackage{amsfonts}
\usepackage{hyperref}
\usepackage{lineno}
\modulolinenumbers[1]
\usepackage{caption}
\usepackage{booktabs}
\usepackage{xcolor}
\usepackage{soul}
\usepackage{tabularx}
\usepackage{rotating}
\journal{arXiv}

\begin{document}

\begin{frontmatter}
%\title{Machine learning-enabled inverse design of thermoelastic metamaterials}
\title{Machine learning-enabled inverse design of bimaterial thermoelastic lattice metamaterials}

\author[tud]{Xiang-Long Peng\corref{cor1}}
\ead{xianglong.peng@tu-darmstadt.de}
\author[tud]{Bai-Xiang Xu}
\cortext[cor1]{Corresponding author}
\address[tud]{Mechanics of Functional Materials Division, Institute of Materials Science, Technische Universität Darmstadt, 64287 Darmstadt, Germany}
\begin{abstract}
The thermoelastic metamaterial based on a bimaterial hybrid-honeycomb structure, exhibiting simultaneously negative Poisson's ratios and negative thermal expansion coefficients is very promising for various application. This work is dedicated to the machine learning (ML)-enabled inverse design of such structure, on the basis of high-throughput simulation and neural network models. A large dataset is generated through computational homogenization of structures with varying geometrical features and base material properties. A forward ML model is first trained to efficiently and accurately predict the effective thermoelastic properties for a given structure design. Subsequently, inverse ML models are developed to suggest geometrical features and base materials for desired target properties. To address various inverse design scenarios, six different models are proposed, each defined by different combinations of target effective properties and structural design variables. The trained forward model is integrated into the loss functions of the inverse models and is also employed to generate additional datasets for cases with fixed base materials. The good predictive performance of the forward and inverse ML models is demonstrated by representative design examples. These ML models can be applied to efficiently solving specific inverse design tasks involved in the practical application of the thermoelastic metamaterial in novel engineering systems.
\end{abstract}


\end{frontmatter}

%\linenumbers
\section{Introduction}
Thermoelastic metamaterials are generally materials exhibiting unconventional effective thermoelastic properties, e.g., negative thermal expansion coefficients. They are usually designed by properly arranging two or more base materials with distinctive thermal expansion coefficients within a periodic porous architecture. Upon heating, they may undergo an overall contraction as a combined result of internal elastic deformations induced by thermal mismatch and the thermal expansion of the base materials, mediated through different mechanisms (e.g., bending-dominated or stretching-dominated \cite{Peng2020novel}). To date, numerous metamaterials with negative thermal expansion coefficeints have been proposed (e.g., \cite{Lakes1996cellular,Sigmund1997design,Grima2007system,Jefferson09,Lim2012negative,Wang2016lightweight,Wei2016planar,Jin17,Xu2018routes,zhang2025windmill} ). Some of them also exhibit negative Poisson’s ratios, i.e., auxetic behavior (e.g., \cite{Ha2015,Ai18,Peng2020novel,Peng2021tunable,tian2023metamaterial,zhang2024series}). They can be fabricated by the rapid developing additive manufacturing technologies and are drawing more and more attentions \cite{dubey2024negative}.

Thermoelastic metamaterials can provide broad tunability of effective properties by tailoring the underlying microstructure and selecting appropriate base materials. This advantage makes them well suited for applications requiring customized thermoelastic responses, such as thermal stress compensation in precision instruments and temperature-adaptive engineering structures. At the same time, it also poses challenges. On the one hand,  forward prediction, i.e, evaluating the effective thermoelastic properties for a given structure design, necessitates establishing the nonlinear correlations between design variables and effective properties. On the other hand, inverse design, i.e., identifying the design variables of a structure that yield specified target properties, is nontrivial and typically demands sophisticated optimization algorithms. 

In this context, rapidly developing machine learning (ML) methods have emerged as promising alternatives to conventional approaches. In recent years, ML techniques have been successfully applied to address challenging problems in the modeling and design of microstructured materials as reviewed in \cite{jin2022intelligent,zheng2023deep,peng2024can,lee2024data}. In particular, their effectiveness in forward prediction and inverse design of metamaterials has been demonstrated in recent studies (e.g., \cite{kulagin2020architectured,fernandez2021anisotropic,bastek2022inverting,pahlavani2022deep,felsch2023controlling,challapalli2023inverse,zheng2023unifying,lin2024machine,cho2024inverse,peng2024data,liang2025demand}). 

In this work, we focus on the inverse design of a thermoelastic lattice metamaterial based on the bimaterial hybrid-honeycomb structure proposed in \cite{Peng2020novel}. This structure, with a simple geometric configuration, simultaneously exhibits negative thermal expansion coefficients, negative Poisson’s ratios, and enhanced stiffness. It is promising for various applications such as structural or functional components in aerospace systems and temperature-adaptive structures \cite{dubey2024negative}. Since its effective thermoelastic properties are governed by the interactions between the structural geometries and base material properties, its inverse design is challenging. Here, we integrate the numerical simulation with ML methods to address this challenge. The computational homogenization method is adopted to calculated the effective thermoelastic properties of a large number of structures with randomly sampled structural features, which yields a large dataset. Forward and inverse ML models based on the artificially neural network (ANN) are constructed, trained, and validated. Their excellent performances are demonstrated by a number of representative forward and inverse design examples.

The remainder of the paper is organized as follows. Section\,\ref{design_model} introduces the bimaterial hybrid-honeycomb structure, the computational homogenization method for evaluating effective thermoelastic constants, and the ML frameworks for forward prediction and inverse design. Section\,\ref{results_discussion} presents data generation, training and validation of the forward and inverse ML models, and their performance demonstration. Finally, Section\,\ref{conclusion} summarizes the work and presents the conclusions.

\section{Design and methods}
\label{design_model}
In this section, the design variables of hybrid-honeycomb structures, the calculation of their effective theromelastic properties, and the ML methods for forward and inverse design are introduced.
\subsection{Design variables of bimaterial hybrid-honeycomb structures}
\begin{figure}[!htbp]
	\centerline{\includegraphics[width=\textwidth]{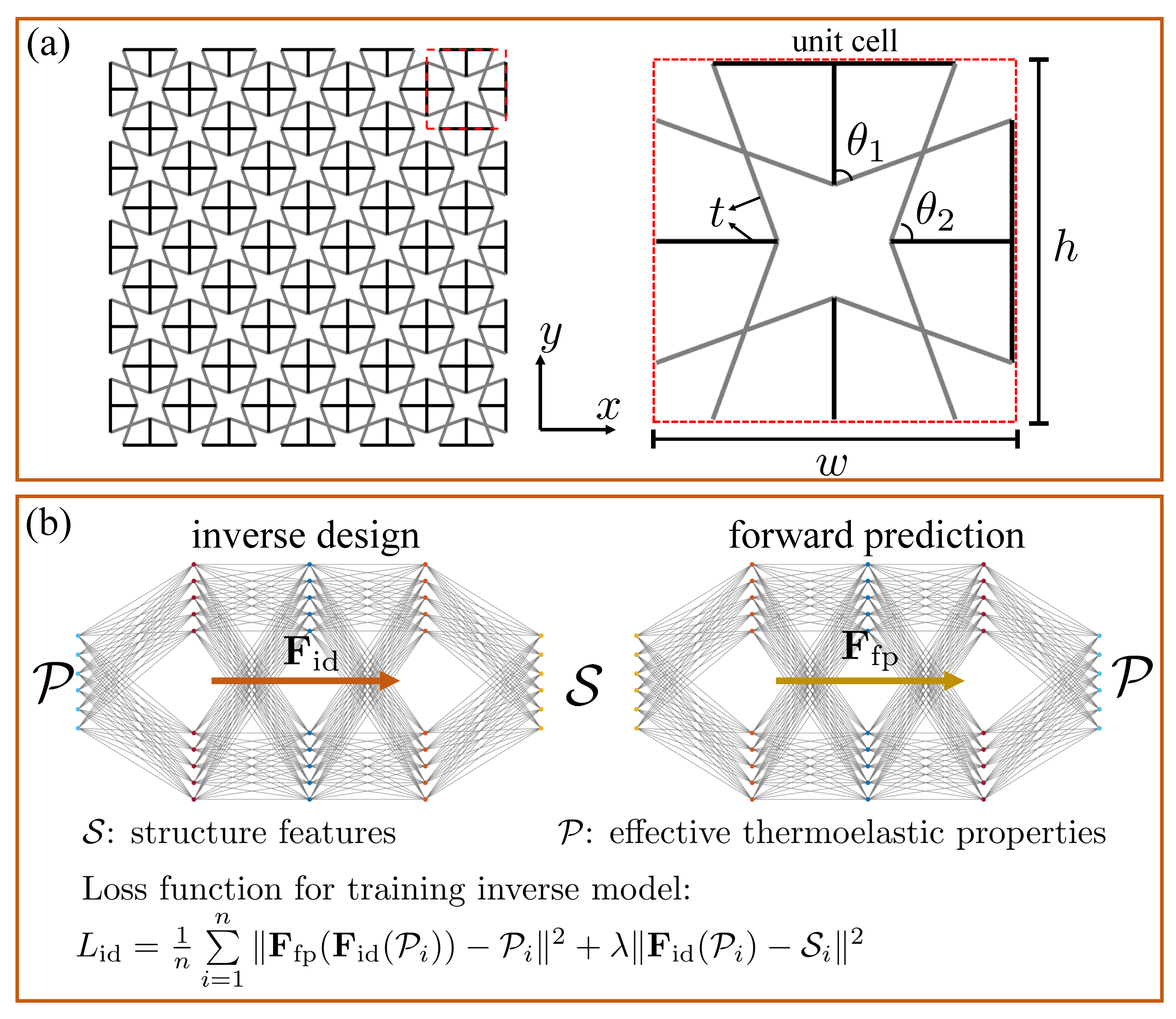}}
	\caption{ML-enabled forward and inverse design of hybrid-honeycomb structure. (a) Geometry of the hybrid-honeycomb structure.  (b) Illustration of ANN models for forward prediction and inverse design. During the training of the inverse ML models, the already trained forward ML model is used to evaluate the loss function. More details are found in Section\,\ref{surrogate_method}. }
	\label{geometry}
\end{figure}

The geometry of the two-dimensional hybrid-honeycomb structure \cite{Peng2020novel} is illustrated in Fig.\,\ref{geometry}a. Its unit cell is characterized by the width $w$, the height $h$, the two inclined angles $\theta_1$ and $\theta_2$, and the strut thickness $t$. Then, the relative density can be expressed as 
\begin{equation}
\rho=\frac{t}{wh}[h[1+\text{cot}\theta_2+\frac{2}{\text{sin}\theta_2}]+w[1+\text{cot}\theta_1+\frac{2}{\text{sin}\theta_1}]].
\end{equation}
Here, we consider the four dominant geometrical features, i.e., the aspect ratio of the unit cell $h/w$, the inclined angles $\theta_1$ and $\theta_2$, and the relative density $\rho$, as the geometrical design variables.

Two base materials with distinct isotropic thermoelastic properties are assigned to the inclined (I) and noninclined struts (N), respectively. The corresponding material parameters are Young's moduli ($E_\text{I}$ and $E_\text{N}$), Poisson's ratios ($\nu_\text{I}$ and $\nu_\text{N}$), and thermal expansion coefficients ($\alpha_\text{I}$ and $\alpha_\text{N}$). The normalized effective thermoelastic properties depend on the Young's modulus ratio $E_\text{I}/E_\text{N}$ and thermal expansion coefficient ratio $\alpha_\text{I}/\alpha_\text{N}$ rather than on their absolute values. Since the influence of Poisson's ratios of the base materials is insignificant, they are assumed to be constant, i.e.,  $\nu_\text{I}= \nu_\text{N}=0.3$ . Therefore, $E_\text{I}/E_\text{N}$ and $\alpha_\text{I}/\alpha_\text{N}$ are chosen as the two base material-related features. Thus, in total, six structural design variables are considered, i.e., $h/w$, $\theta_1$, $\theta_2$, $\rho$, $E_\text{I}/E_\text{N}$, and $\alpha_\text{I}/\alpha_\text{N}$. Their design ranges are defined in Table 1.

\begin{table}[htbp]
	\caption{The ranges of the six design variables are summarized. To avoid unfeasible geometries, the ranges of the two strut angles are shift by $5^\circ$ from their theoretical lower and upper bounds. }
	{\small
		%\begin{tabular}{cc}
		\begin{tabular}{p{0.5\textwidth}p{0.5\textwidth}}
			\toprule
			\small
			Design variables & Range\\
			\midrule
			$E_\text{I}/E_\text{N}$ & $[10^{-3},\,10^{3}]$ \\
			 $\alpha_\text{I}/\alpha_\text{N}$  & $[10^{-3},\,10^{0}]$\\
			$\rho$ & $[0.1,\,0.5]$ \\
			$h/w$& $[1/3,\,3]$\\
			$\theta_1$ &$[\text{arctan}(\frac{w}{h})+5^\circ,\,175^\circ-\text{arctan}(\frac{w}{h})]$\\
		  $\theta_2$& $[\text{arctan}(\frac{h}{w})+5^\circ,\,175^\circ-\text{arctan}(\frac{h}{w})]$\\ 
			\bottomrule
	\end{tabular}}
	\label{table_cr}
\end{table}
\subsection{Computational homogenization of effective thermoelastic properties}
\label{Comp_homo}
Macroscopically, the hybrid-honeycomb structures are effectively treated as homogeneous materials. Considering the geometrical symmetry of the unit cell, in the selected coordinate system (see Fig.\,\ref{geometry}a), the corresponding two-dimensional effective linear thermoelastic constitutive relation is expressed in Voigt notation as follows,
\begin{equation}
	\begin{bmatrix}
		\sigma^{\text{M}}_{xx}
		\\
		\sigma^{\text{M}}_{yy}
		\\
		\sigma^{\text{M}}_{xy}
	\end{bmatrix}
	=\begin{bmatrix}
		\frac{1}{E^{\star}_{x}} & -\frac{\nu^{\star}_{yx}}{E^{\star}_{y}} & 0\\
		-\frac{\nu^{\star}_{xy}}{E^{\star}_{x}} &  \frac{1}{E^{\star}_{y}}&0\\
		0 & 0&\frac{1}{G^{\star}_{xy}}
	\end{bmatrix}^{-1}\begin{bmatrix}
		\epsilon^{\text{M}}_{xx}-\alpha_{x}^\star\Delta T
		\\
		\epsilon^{\text{M}}_{yy}-\alpha_{y}^\star\Delta T
		\\
		2\epsilon^{\text{M}}_{xy}\end{bmatrix}.
\end{equation}
where $\sigma^{\text{M}}_{ij}$ and $\epsilon^{\text{M}}_{ij}$ ($i,\,j=x,\,y$) denote the macroscopic stress and strain components which are defined as the volume averages of their microscopic counterparts, and $\Delta T$ is the temperature increase. $E^{\star}_{i}$, $\nu^{\star}_{ij}$, $G^{\star}_{ij}$, and $\alpha_{i}^\star$ are the effective Young's modulus, Poisson's ratio, shear modulus, and thermal expansion coefficient, respectively. The symmetry of the effective stiffness matrix yields $-\nu^{\star}_{yx}/E^{\star}_{y}=-\nu^{\star}_{xy}/E^{\star}_{x}$. Thus, there are 6 independent effective thermoelastic constants, i.e., $E^{\star}_{x}$, $E^{\star}_{y}$, $\nu^{\star}_{xy}$, $G^{\star}_{yx}$, $\alpha_{x}^\star$, and $\alpha_{y}^\star$.

For a given structure design, the effective thermoelastic constants are calculated by finite element (FE) simulations based on the computational homogenization method. The unit cells are taken as the representative volume elements (RVEs). Each strut is meshed by one Timoshenko beam element. In each case, two axial loading and one shearing loading cases by prescribing macroscopic strain components are simulated to determine the resulting macroscopic stress components. Each loading case yields one column in the stiffness matrix. In addition, one thermal loading by prescribing a constant temperature increase is simulated to determine the effective thermal expansion coefficients. Periodic boundary conditions are exploited. The FE simulations are conducted with in-house Matlab codes.
\subsection{ML models for forward prediction and inverse design}
\label{surrogate_method}
\subsubsection{Forward prediction}
\label{forward_ML}
As illustrated in Fig.\,\ref{geometry}b, the ML model for forward prediction $\textbf{F}_\text{fp}$ takes the structural features $\mathcal{S}$ of a design as the inputs and predict the corresponding effective thermoelastic properties $\mathcal{P}$. Here, the conventional ANN model is employed for this purpose. As discussed previously, there are 6 structural features and 6 effective thermoelastic constants, i.e.,
\begin{equation}
	\begin{split}
	\mathcal{S}&=[E_\text{I}/E_\text{N},\, \alpha_\text{I}/\alpha_\text{N},\, \rho,\, h/w,\, \theta_1, \, \theta_2]^\text{T},\\
	\mathcal{P}&=[E^{\star}_{x}/E_\text{N},\,E^{\star}_{y}/E_\text{N},\, \nu^{\star}_{xy},\, G^{\star}_{yx}/E_\text{N},\, \alpha_{x}^\star/\alpha_\text{N},\, \alpha_{y}^\star/\alpha_\text{N}]^\text{T}.
	\end{split}
\end{equation}
Note that the above structural features and the effective thermoelastic constants are scaled prior to model training. The structural features and the effective properties are normalized by min-max scaling to the ranges $[0,\,1]$ and $[-1,\,1]$, respectively. Before normalization, the features $E_\text{I}/E_\text{N}$ and $\alpha_\text{I}/\alpha_\text{N}$ and the effective moduli $E^{\star}_{x}/E_\text{N}$, $E^{\star}_{y}/E_\text{N}$, and $G^{\star}_{yx}/E_\text{N}$ are first transformed using the base-10 logarithm. 

The forward ML model is trained towards minimizing a defined loss function $L_\text{fp}$ by optimizing the model parameters (i.e., the weights and bias). Here, $L_\text{fp}$ quantifies the mean square error between the effective properties predicted by the ML model, i.e., $\textbf{F}_\text{fp}(\mathcal{S})$ and the corresponding ground truth $\mathcal{P}$, i.e.,
\begin{equation}
	L_\text{fp}=\frac{1}{n}\sum\limits_{i=1}^{n}\| \textbf{F}_\text{fp}(\mathcal{S}_i)-\mathcal{P}_i \|^2,
\end{equation}
where $\mathcal{S}_i$ and $\mathcal{P}_i$ are the structural features and the effective properties of the $i$th data point, respectively, and $n$ denotes the batch size, i.e., the number of data points used in each iteration. 

\subsubsection{Inverse design}
\label{inverse_ML}
The inverse design refers to predict a structural design with target effect properties. In general, there are two types of inverse design methods, i.e., indirect and direct methods (e.g., \cite{peng2024can,otto2025data}). The former usually involves an iterative searching process via different algorithms such as the genetic algorithm (e.g., \cite{maurizi2022inverse}) and Bayesian optimization (e.g., \cite{hu2023multi}). In this work, we employ the direct method. In this case, an ML model $\textbf{F}_\text{id}$ which directly predicts the structural features $\mathcal{S}$ for given target effective properties $\mathcal{P}$ is to be constructed. To resolve the nonuniqueness issue that multiple designs may result in the same effective properties, the following loss function $L_\text{id}$ is introduced (e.g., \cite{bastek2022inverting,peng2024data,liu2018training}),
\begin{equation}
	L_\text{id}=\frac{1}{n}\sum\limits_{i=1}^{n}\| \textbf{F}_\text{fp}(\textbf{F}_\text{id}(\mathcal{P}_i))-\mathcal{P}_i \|^2+\lambda\| \textbf{F}_\text{id}(\mathcal{P}_i)-\mathcal{S}_i \|^2,
	\label{L_id}
\end{equation}
In the above equation, the first term represents the reconstruction loss, quantifying the difference between the effective properties of the designed structure (evaluated by the trained forward ML model, i.e., $\textbf{F}_\text{fp}(\textbf{F}_\text{id}(\mathcal{P}_i))$) and the target values. The second term captures the prediction loss scaled by a factor $\lambda$, i.e., the discrepancy between the predicted structural features and those associated with the inputs. This term is only involved (by $\lambda=0.2$) at the beginning of training to promote convergence and is then dropped (by $\lambda=0$). Thus, the inverse ML model is essentially trained towards minimizing the reconstruction loss. The ANN models are selected as the inverse ML models. Several different inverse ML models with different number of inputs and outputs will be constructed for different inverse design tasks, more details will be given in the relevant sections. The current inverse design model predicts one certain structure design for a given set of target properties, which is well-suited for the lattice structures characterized by a few defined features. For the  inverse design of sophistic microstructures, one may employ generative ML models to design multiple microstructures with the same target properties (e.g., \cite{wang2020deep, zheng2021controllable}).

The deep learning framework Pytorch \cite{paszke2019pytorch} is exploited to construct and train forward and inverse ML models. We iteratively test various combinations of hyperparameters including the ANN architecture, the learning rate, the epoch number, and the batch size  to attain ML models with satisfactory performance. 

\section{Results and discussion}
\label{results_discussion}
In this section, we first introduce the data generation. Then, the training and evaluation of the forward prediction ML model are discussed. Afterwards, a number of inverse design ML models are trained and evaluated. Some representative examples to demonstrate the performance of the forward and inverse ML models are illustrated. Some of the relevant figures and tables involved in the discussion are found in the supplementary material and are labeled with an 'S' followed by the corresponding number.
\subsection{Dataset generation and property range}
\label{dataset}
\begin{figure}[!htbp]
	\centerline{\includegraphics[width=\textwidth]{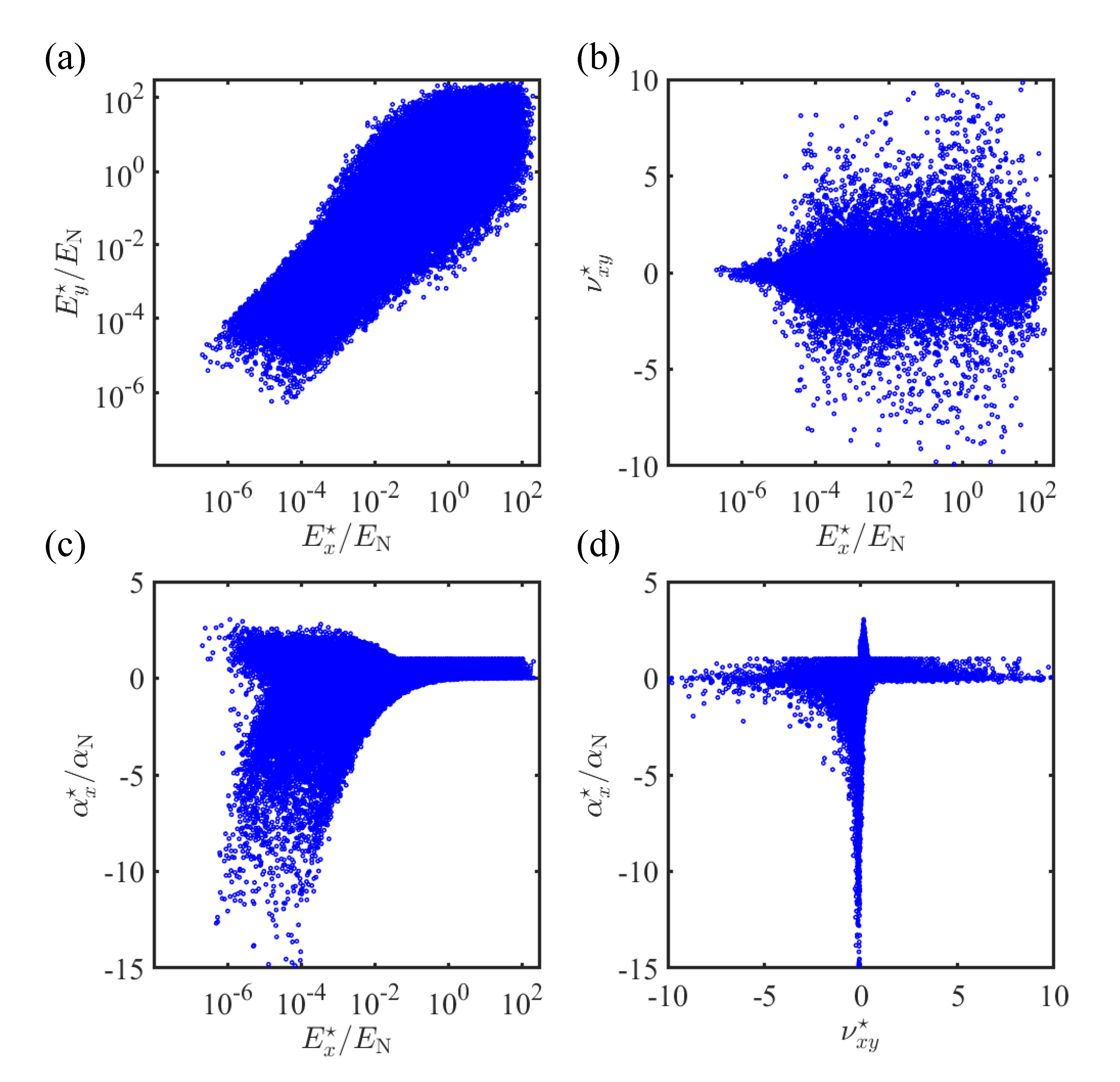}}
	\caption{Effective thermoelastic properties of structures in the dataset:  (a) $E^{\star}_{x}/E_\text{N}$ versus $E^{\star}_{y}/E_\text{N}$, (b) $E^{\star}_{x}/E_\text{N}$ versus $\nu^{\star}_{xy}$, (c) $E^{\star}_{x}/E_\text{N}$ versus $\alpha^{\star}_{x}/\alpha_\text{N}$, and (d) $\nu^{\star}_{xy}$ versus $\alpha^{\star}_{x}/\alpha_\text{N}$. The effective properties vary broadly in the property space. The dataset is representative. }
	\label{data}
\end{figure}

To obtain a representative dataset, $10^5$ structures are designed by randomly sampling the six design variables within their defined ranges in Table\,\ref{table_cr}. The effective thermoelastic properties of these structures are calculated by the computational homogenization method introduced in Section\,\ref{Comp_homo}. Thus, in the dataset of $10^5$ data points, each data point consists of the structural features of a structure paired with its effective thermoelastic properties. As demonstrated by the plots of $E^{\star}_{x}/E_\text{N}$ versus $E^{\star}_{y}/E_\text{N}$, $E^{\star}_{x}/E_\text{N}$ versus $\nu^{\star}_{xy}$,  $E^{\star}_{x}/E_\text{N}$ versus $\alpha^{\star}_{x}/\alpha_\text{N}$, and $\nu^{\star}_{xy}$ versus $\alpha^{\star}_{x}/\alpha_\text{N}$ in Fig.\,\ref{data}, the effective properties of the hybrid-honeycomb lattice structures are tunable in broad ranges. The effective Young's moduli and shear modulus span several orders of magnitude. Both the effective Poisson’s ratio and thermal expansion coefficients exhibit a continuous transition from large positive to small negative values. The data points are broadly distributed across the property space, indicating that the dataset is representative. The dataset is accessible at \url{https://github.com/XiangLongPeng/ML_inverse_design_thermo_metamater.git}. 80\% of the data points in the dataset (Section \ref{dataset}) are used for training, and the remaining 20\% are used for testing.
\subsection{ML-enabled forward prediction}
\label{forward_model1}
\begin{figure}[!htbp]
	\centerline{\includegraphics[width=\textwidth]{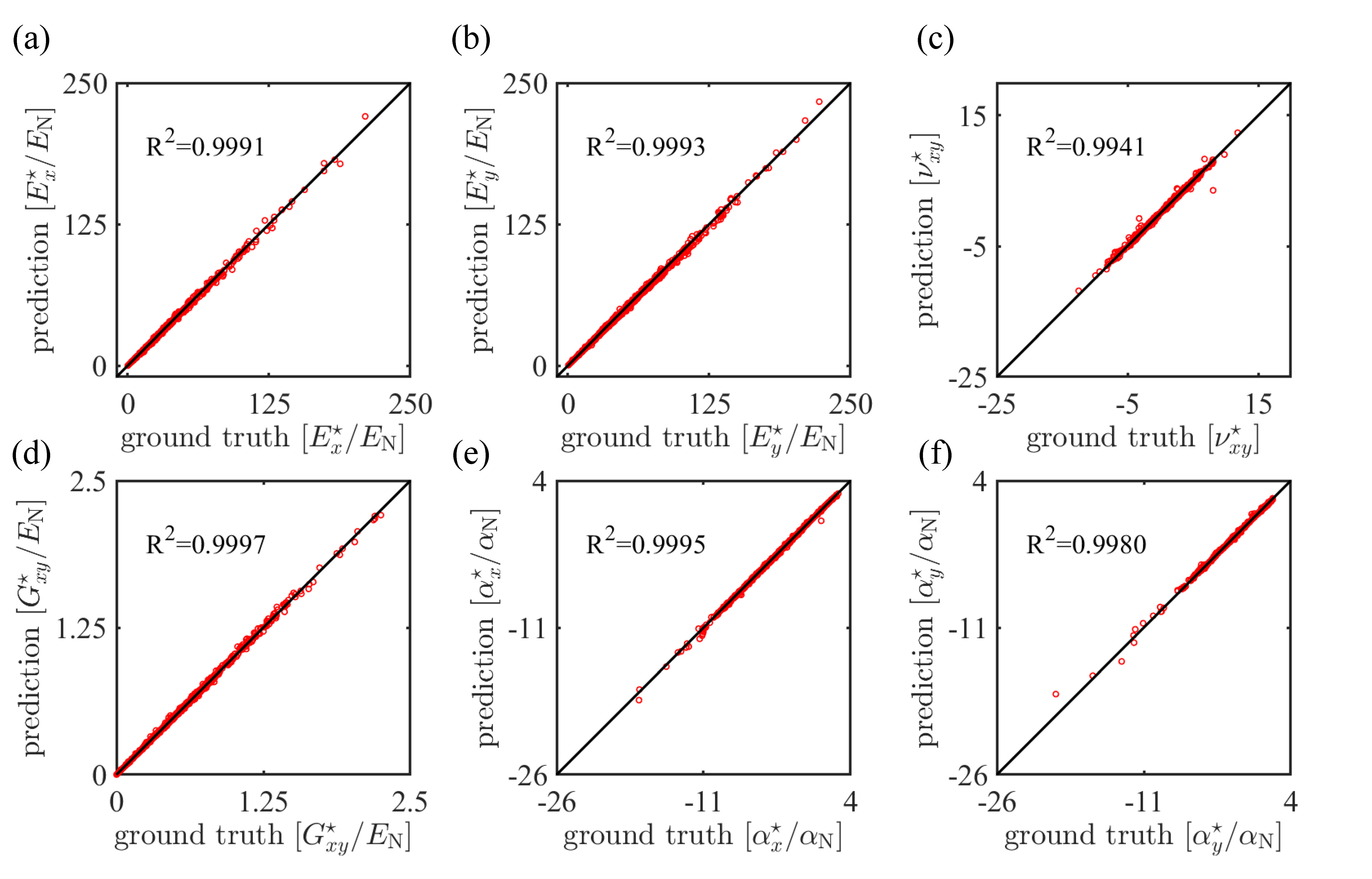}}
	\caption{Comparison between forward ML model prediction and the ground truth from FE simulations: (a) $E^{\star}_{x}/E_\text{N}$, (b) $E^{\star}_{y}/E_\text{N}$, (c) $\nu^{\star}_{xy}$, (d) $G^{\star}_{xy}/E_\text{N}$, (e) $\alpha^{\star}_{x}/\alpha_\text{N}$, and (f) $\alpha^{\star}_{y}/\alpha_\text{N}$. The ML model prediction shows good agreement with the ground truth with a coefficient of determination (i.e., $\text{R}^2$) higher than 0.994 for all six properties. }
	\label{R_square_forward}
\end{figure}

\begin{figure}[!htbp]
	\centerline{\includegraphics[width=\textwidth]{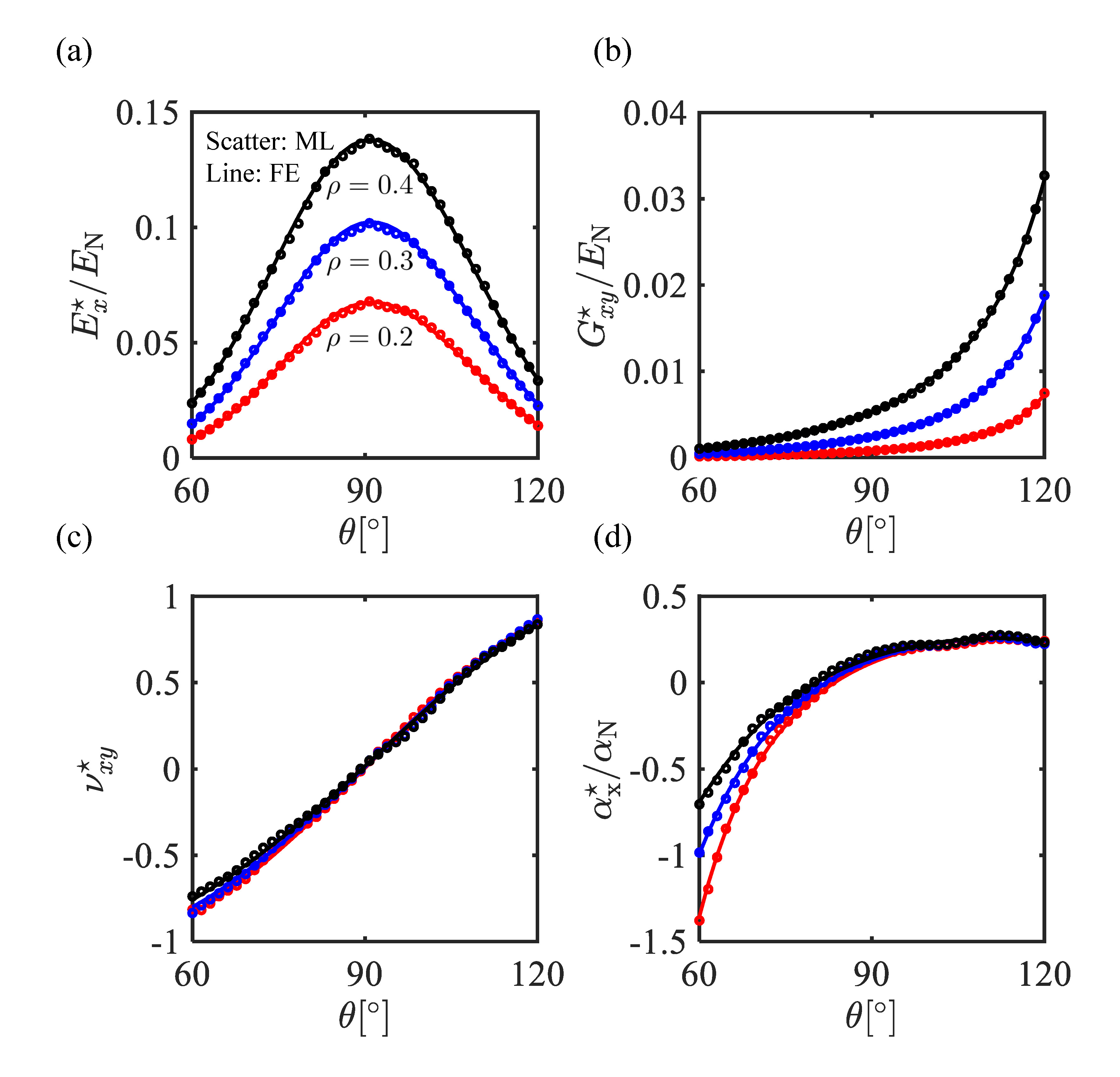}}
	\caption{Effective properties varying with the strut angle $\theta_1=\theta_2=\theta$ and the relative density $\rho$: (a) $E^{\star}_{x}/E_\text{N}$,  (b) $G^{\star}_{xy}/E_\text{N}$, (c) $\nu^{\star}_{xy}$, and (d) $\alpha^{\star}_{x}/\alpha_\text{N}$. The ML model prediction aligns well with those from FE simulations.}
	\label{G_P_curves}
\end{figure}

The ML model for forward prediction is constructed and trained based on the method described in Section\,\ref{forward_ML}.  The finally used hyperparameters are listed in Table S.1 in the supplementary material. According to the training and test loss-epoch curves in Fig. S.1 in the supplementary material, the forward model is properly trained without underfitting or overfitting. The comparison between the effective properties predicted by ML model and those from FE simulations (i.e., the ground truth) for structures in the test dataset is shown in Fig.\,\ref{R_square_forward}. The high coefficients of determination (i.e., $\text{R}^2$ factors larger than 0.994) for all six properties indicate that the ML prediction matches the ground truth well, which demonstrates the good performance of the forward model. In Fig.\,\ref{G_P_curves},  the effective properties as functions of the strut angle $\theta_1=\theta_2=\theta$ and the relative density $\rho$ predicted by the ML model are plotted, which agrees well with the simulation results. These together demonstrate the good performance of the forward prediction model. This ML model can act as an efficient property predictor for the practical application of the hybrid-honeycomb structure. 

\subsection{ML-enabled inverse design}
The inverse design refers to designing a structure with target effective properties. In practice, the six effective thermoelastic constants can be either fully or partially specified as design targets. Some design variables may also be fixed. For instance, the base material properties are fixed once the two base materials for fabrication are selected. Thus, we construct and train six inverse ML models with different combinations of inputs and outputs (see Table \ref{inverse_in_out}), which are categorized into four cases. The inverse models are constructed and trained by employing the method described in Section\,\ref{inverse_ML}. The finally chosen hyperparameters are listed in Table\,S.1. The loss-epoch curves for the training of the six inverse design ML models are plotted in Fig.\,S.2 in the supplementary material, which indicate that they are properly trained without overfitting or underfitting.
\begin{table}[htbp]
	\caption{The inputs and outputs of the six inverse ML models. }
	{\small
			\begin{tabular}{p{0.1\textwidth}p{0.3\textwidth}p{0.3\textwidth}p{0.3\textwidth}}
			\toprule
			Inverse model No. & Inputs \newline(effective properties) & Outputs \newline(structural features)&Comments\\
			\midrule		
	  1 & $E^{\star}_{x}/E_\text{N}$,\,$E^{\star}_{y}/E_\text{N}$,\, $\nu^{\star}_{xy}$,\newline $G^{\star}_{yx}/E_\text{N}$,\,$\alpha_{x}^\star/\alpha_\text{N}$,\, $\alpha_{y}^\star/\alpha_\text{N}$ & $E_\text{I}/E_\text{N}$,\, $\alpha_\text{I}/\alpha_\text{N}$,\, $\rho$,\, $h/w$,\, $\theta_1$, \, $\theta_2$& See Section\,\ref{Case1}\\
	  2 & $\nu^{\star}_{xy}$,\, $\alpha_{x}^\star/\alpha_\text{N}$& $E_\text{I}/E_\text{N}$,\, $\alpha_\text{I}/\alpha_\text{N}$,\, $\rho$,\, $h/w$,\, $\theta_1$, \, $\theta_2$& See Section\,\ref{Case2}\\
	  3 & $E^{\star}_{x}/E_\text{N}$,\,$\alpha_{x}^\star/\alpha_\text{N}$& $E_\text{I}/E_\text{N}$,\, $\alpha_\text{I}/\alpha_\text{N}$,\, $\rho$,\, $h/w$,\, $\theta_1$, \, $\theta_2$ &See Section\,\ref{Case2}\\
	  4 & $E^{\star}_{x}/E_\text{N}$,\,$E^{\star}_{y}/E_\text{N}$,\, $\nu^{\star}_{xy}$,\newline $G^{\star}_{yx}/E_\text{N}$,\,$\alpha_{x}^\star/\alpha_\text{N}$,\, $\alpha_{y}^\star/\alpha_\text{N}$& $\rho$,\, $h/w$,\, $\theta_1$, \, $\theta_2$ & See Section\,\ref{Case3}; applicable for specified $E_\text{I}/E_\text{N}$ and  $\alpha_\text{I}/\alpha_\text{N}$.\\
	  5 & $\nu^{\star}_{xy}$,\, $\alpha_{x}^\star/\alpha_\text{N}$ & $\rho$,\, $h/w$,\, $\theta_1$, \, $\theta_2$&See Section\,\ref{Case4}; applicable for specified $E_\text{I}/E_\text{N}$ and  $\alpha_\text{I}/\alpha_\text{N}$\\
	  6 & $E^{\star}_{x}/E_\text{N}$,\,$\alpha_{x}^\star/\alpha_\text{N}$& $\rho$,\, $h/w$,\, $\theta_1$, \, $\theta_2$ &See Section\,\ref{Case4}; applicable for specified $E_\text{I}/E_\text{N}$ and $\alpha_\text{I}/\alpha_\text{N}$\\
			\bottomrule
		\end{tabular}
	}\label{inverse_in_out}
\end{table}

\subsection{Case 1: full design variables and full target properties}
\label{Case1}
In this case, all six properties act as the inputs (i.e., design targets) and the six design variables are the outputs for inverse ML model 1. This model is trained with the dataset in Section\,\ref{dataset}. As shown in Fig.\,S.4 in the supplementary material, the six effective thermoelastic constants of inversely designed structures align well with the target ones from the test dataset. 

To further demonstrate the performance of the inverse model, three representative design examples are considered. Since the effective properties of the hybrid-honeycomb structure are confined within certain ranges, a corresponding structure may not exist for arbitrarily chosen target properties. Therefore, in the first two examples, the target properties are taken from two known structures. As shown in Table \ref{inverse1}, the inverse model successfully predicts structures with effective properties matching the target properties with only small relative errors. Moreover, the inversely designed structures differ in their structural features from those used to generate the target properties. This indicates that multiple distinct structures can yield the same effective properties and that the inverse model is capable of identifying one such structure. In the third example, the target properties are specified by perturbing those in the first example, and the inverse model again predicts a structure that matches the target properties. These examples demonstrate that the inverse model can successfully design structures for specified effective properties provided that they lie within the achievable range.

\begin{table}[htbp]
	\caption{Inverse design results obtained using inverse ML model 1. In the first example, the target properties correspond to a known structure with $E_\text{I}/E_\text{N}=1$, $\alpha_\text{I}/\alpha_\text{N}=0.1$, $\rho=0.2$, $h/w=1$, $\theta_1=60^\circ$, and $\theta_2=60^\circ$. In the second example, they correspond to a known structure with $E_\text{I}/E_\text{N}=0.1$, $\alpha_\text{I}/\alpha_\text{N}=0.01$, $\rho=0.3$, $h/w=0.8$, $\theta_1=70^\circ$, and $\theta_2=60^\circ$. In the third example, the target properties are obtained by perturbing those in the first example. }
	{\small
		\begin{tabular}{p{0.3\textwidth}p{0.3\textwidth}p{0.3\textwidth}p{0.1\textwidth}}
			%\begin{tabular}{lllll}
			\toprule
			Target properties: \newline$E^{\star}_{x}/E_\text{N}$,\,$E^{\star}_{y}/E_\text{N}$,\, $\nu^{\star}_{xy}$,\newline $G^{\star}_{yx}/E_\text{N}$,\,$\alpha_{x}^\star/\alpha_\text{N}$,\, $\alpha_{y}^\star/\alpha_\text{N}$ & Designed properties:\newline$E^{\star}_{x}/E_\text{N}$,\,$E^{\star}_{y}/E_\text{N}$,\, $\nu^{\star}_{xy}$,\newline $G^{\star}_{yx}/E_\text{N}$,\,$\alpha_{x}^\star/\alpha_\text{N}$,\, $\alpha_{y}^\star/\alpha_\text{N}$&Designed features: \newline $E_\text{I}/E_\text{N}$,\, $\alpha_\text{I}/\alpha_\text{N}$,\newline $\rho$,\, $h/w$,\, $\theta_1$, \, $\theta_2$ & Relative error\\
			\midrule
			0.00816,\,0.00816,\newline-0.835,\,0.000135,\newline-1.355,\,-1.355 &0.00831,\,0.00772,\newline-0.846,\,0.000139,\newline-1.284,\,-1.356 &0.71,\,0.061,\newline0.21,\,1.00,\newline$62.0^\circ$,\,$60.3^\circ$
			&3.4\%\\ \\
			0.00459,\,0.00241,\newline-1.006,\,0.000136,\newline-0.638,\,-1.309 &0.00481,\,0.00234,\newline-1.102,\,0.000129,\newline-0.604,\,-1.355 &0.13,\,0.11,\newline0.29,\,0.85,\newline$69.8^\circ$,\,$56.1^\circ$
			&6.3\%\\ \\
			0.01,\,0.01,\newline-0.8,\,0.0002,\newline-1,\,-1 &0.01,\,0.0097,\newline-0.8,\,0.0002,\newline-0.899,\,-0.976 &0.54,\,0.10,\newline0.24,\,1.01,\newline$64.6^\circ$,\,$62.1^\circ$
			&6.4\%\\
			\bottomrule
		\end{tabular}
	}
	\label{inverse1}
\end{table}
\subsection{Case 2: full design variables and partial target properties}
\label{Case2}
In this case, only two of the six effective thermoelastic constants are taken as the target properties and the six structural features are considered as design variables. In inverse ML model 2, the effective Poisson's ratio $\nu^{\star}_{xy}$ and thermal expansion coefficient $\alpha_{x}^\star/\alpha_\text{N}$ are the target properties. In inverse ML model 3, the effective Young's modulus$E^{\star}_{x}/E_\text{N}$ and thermal expansion coefficient $\alpha_{x}^\star/\alpha_\text{N}$ are considered. The two inverse models are trained with the dataset in Section\,\ref{dataset} by extracting the relevant two properties of each data point. Figs.\,S.4 and S.5 show that both inverse models can predict structures with target properties from the test dataset with high coefficients of determination, indicating their good performance.

A few representative inverse design examples by using inverse model 2 are presented in Table\,\ref{inverse2}. The corresponding results for inverse model 3 are found in Table\,S.2.
Since only two thermoelastic properties are specified, it is more likely to find out a structure matching the targets compared to case 1. Thus, the target properties can be directly specified in the design examples guided by the property range in Fig.\,\ref{data}d. In the first example, the target properties are $\nu^{\star}_{xy}=-1.0$ and $\alpha_{x}^\star/\alpha_\text{N}=-1.0$. Taking these two values as inputs, the inverse model predicts the corresponding structure with structural features listed in Table\,\ref{inverse2}. The relative error between the target and designed properties is only 2.5\%. 

Then, the inverse model is used to determine the minimum and maximum of the attainable $\alpha_{x}^\star/\alpha_\text{N}$ for a specified $\nu^{\star}_{xy}$ and, conversely, the minimum and maximum of $\nu^{\star}_{xy}$ for a given $\alpha_{x}^\star/\alpha_\text{N}$. These are realized by the following iterative search method. To find the minimum of a property, the search starts at an initial guess equal to the minimum value of this property in the dataset, which is then increased step by step. At each step, the inverse model takes the candidate value of this property and the other specified property as inputs and predict the corresponding structural features. The process terminates once a structure is found that satisfies the target properties with relative error less than 5\%, and the corresponding value is taken as the minimum of the property. Conversely, to find the maximum of a property, the search starts at a value being the maximum value of this property in the dataset, which is then decreased step by step until a feasible structure is identified. The obtained value is then the maximum. As listed in Table\,\ref{inverse2}, by employing this method, the minimum and maximum of $\alpha_{x}^\star/\alpha_\text{N}$ for $\nu^{\star}_{xy}=-1.0$, and those of $\nu^{\star}_{xy}$ for $\alpha_{x}^\star/\alpha_\text{N}=-1.0$ are successfully determined. The corresponding results for inverse model 3 are found in Table\, S.2 in the supplementary material.

The above two inverse models are particularly useful when only a subset of effective properties is of interest. They enable efficient prediction of structures with specified target properties and can also tackle constrained optimization tasks, i.e., determining the attainable minimum and maximum values of one property when the other is fixed.
\begin{table}[htbp]
	\caption{Inverse design results obtained using inverse ML model 2. In the first design example, the inverse model is used to design the structure with specified target properties  $\nu^{\star}_{xy}$ and $\alpha_{x}^\star/\alpha_\text{N}$. In the other four examples, the results show the minimum and maximum values of one of the two properties when the other is specified.}
	{\small
		\begin{tabular}{p{0.25\textwidth}p{0.25\textwidth}p{0.3\textwidth}p{0.2\textwidth}}
			%\begin{tabular}{lllll}
			\toprule
			Target properties: \newline$\nu^{\star}_{xy}$,\,$\alpha_{x}^\star/\alpha_\text{N}$ & Designed properties:\newline$\nu^{\star}_{xy}$,\,$\alpha_{x}^\star/\alpha_\text{N}$&Designed features: \newline $E_\text{I}/E_\text{N}$,\, $\alpha_\text{I}/\alpha_\text{N}$,\newline $\rho$,\, $h/w$,\, $\theta_1$, \, $\theta_2$ & Relative error\\
			\midrule
			-1.0,\,-1.0 &-1.024,\,-1.025 &0.0638,\,0.00736,\newline0.243,\,0.871,\newline$67.0^\circ$,\,$58.1^\circ$
			&2.5\%\\ 
			-1.0,\, minimum &-0.955,\,-2.738 &0.0105,\,0.00291,\newline0.195,\,0.992,\newline$57.0^\circ$,\,$55.0^\circ$
			&4.5\% \newline(w.r.t. $\nu^{\star}_{xy}$)\\	
		-1.0,\, maximum &-1.04,\,0.68 &0.703,\,0.747,\newline0.245,\,0.880,\newline$81.0^\circ$,\,$55.7^\circ$
			&4\%\newline(w.r.t. $\nu^{\star}_{xy}$)\\	
			minimum,\, -1.0 &-7.58,\,-0.975 &0.0215,\,0.0181,\newline0.170,\,0.347,\newline$80.2^\circ$,\,$24.4^\circ$
			&2.5\%\newline(w.r.t. $\alpha_{x}^\star/\alpha_\text{N}$)\\	
			maximum,\, -1.0 &0.0388,\,-0.976&2.1,\,0.00553,\newline0.260,\,1.686,\newline$55.6^\circ$,\,$106.8^\circ$
			&2.4\%\newline(w.r.t. $\alpha_{x}^\star/\alpha_\text{N}$)\\		
			\bottomrule
		\end{tabular}
	}
	\label{inverse2}
\end{table}
\subsection{Case 3: partial design variables and full target properties}
\label{Case3}
In Cases 1 and 2, both base material-related features and geometrical features are treated as design variables. In practice, the available base materials for fabricating the lattice structures are usually limited to certain types. Thus, in Case 3 presented here and in Case 4 discussed in the following subsection, the two material-related structural features are prescribed, while the four geometrical variables act as the design variables. Without loss of generality, two metallic materials AL 6061 and Invar-36 which are used in \cite{Xu2018routes} to fabricate thermal metamaterials, act as the base materials for the non-inclined and inclined struts, respectively. Accordingly, the two material-related features are $E_\text{I}/E_\text{N}=1.977$ and $\alpha_\text{I}/\alpha_\text{N}=0.0652$ (see \cite{Xu2018routes}). 

The dataset in Section\,\ref{dataset} is not applicable to the present case since the material-related features are randomly sampled therein. Therefore, a new dataset is generated by specifying the material features as defined above and randomly sampling the four geometrical variables. The data generation is performed by exploiting the accurate forward prediction ML model in Section\,\ref{forward_model1} to evaluate the effective properties of generated structures. The total number of data points are 20000, which is much smaller than that of the first dataset but sufficient, given that only four design variables are involved. This new dataset is visualized in Fig.\,S.3. in the supplementary material. It is divided into a training dataset (80\%) and a test dataset (20\%), which are used to train and test inverse ML models 4 here and inverse ML models 5 and 6 in Case 4.

In the present case, the six properties are taken as design targets. Thus, inverse ML model 4 consists of 6 inputs and 4 outputs  (see Table \ref{inverse_in_out}).  Fig.\,S.7 in the supplementary material shows the good agreement between the effective thermoelastic constants of inversely designed structures and the target ones in the test dataset. 

Similar to Case 1, three representative inverse design examples are presented in Table\,\ref{inverse4}. The target properties in the first two inverse tasks correspond to those of two known structures, while in the third example, the target properties are obtained by perturbing those in the first example. In all three examples, inverse ML mode 4 successfully predicts a structure whose effective properties closely math the target ones, demonstrating its good performance.
\begin{table}[htbp]
	\caption{Inverse design results obtained using inverse ML model 4. In the first two examples, target properties are from known structures. In the first example, the corresponding structure features are $\rho=0.2$, $h/w=1$, $\theta_1=60^\circ$, and $\theta_2=60^\circ$. In the second example, they are $\rho=0.3$, $h/w=0.8$, $\theta_1=70^\circ$, and $\theta_2=60^\circ$. In the third example, the target properties are determined by perturbing those in the first example. The inverse model predicts a structure with effective properties matching the target ones in each case.}
	{\small
		\begin{tabular}{p{0.3\textwidth}p{0.3\textwidth}p{0.3\textwidth}p{0.1\textwidth}}
			%\begin{tabular}{lllll}
			\toprule
			Target properties: \newline$E^{\star}_{x}/E_\text{N}$,\,$E^{\star}_{y}/E_\text{N}$,\, $\nu^{\star}_{xy}$,\newline $G^{\star}_{yx}/E_\text{N}$,\,$\alpha_{x}^\star/\alpha_\text{N}$,\, $\alpha_{y}^\star/\alpha_\text{N}$ & Designed properties:\newline$E^{\star}_{x}/E_\text{N}$,\,$E^{\star}_{y}/E_\text{N}$,\, $\nu^{\star}_{xy}$,\newline $G^{\star}_{yx}/E_\text{N}$,\,$\alpha_{x}^\star/\alpha_\text{N}$,\, $\alpha_{y}^\star/\alpha_\text{N}$&Designed features: \newline $\rho$,\, $h/w$,\, $\theta_1$, \, $\theta_2$ & Relative error\\
			\midrule
			0.0128,\,0.0128,\newline-0.818,\,0.000203,\newline-1.281,\,-1.281 &0.0127,\,0.0121,\newline-0.840,\,0.000190,\newline-1.285,\,-1.317 &0.196,\,1.072,\newline$61.7^\circ$,\,$58.3^\circ$
			&2.1\%\\ \\
			0.0701,\,0.0288,\newline-1.018,\,0.000893,\newline-0.390,\,-0.822 &0.0694,\,0.0305,\newline-0.984,\,0.000997,\newline-0.395,\,-0.787 &0.315,\,0.914,\newline$71.0^\circ$,\,$58.0^\circ$
			&3.6\%\\ \\
			0.015,\,0.015,\newline-0.8,\,0.00015,\newline-1.0,\,-1.0 &0.0158,\,0.0151,\newline-0.789,\,0.000137,\newline-1.06,\,-1.09 &0.165,\,1.090,\newline$65.0^\circ$,\,$61.3^\circ$
            &6.5\%\\ 
			\bottomrule
		\end{tabular}
	}
	\label{inverse4}
\end{table}

\subsection{Case 4: partial design variables and partial target properties}
\label{Case4}
Similar to Case 3 in Section\,\ref{Case3},  the effective properties are partially specified as design targets in the present case.  Inverse ML model 5 takes the effective Poisson's ratio $\nu^{\star}_{xy}$ and thermal expansion coefficient $\alpha_{x}^\star/\alpha_\text{N}$ as the inputs. Inverse ML model 6 is designed for the effective Young's modulus $E^{\star}_{x}/E_\text{N}$ and the thermal expansion coefficient $\alpha_{x}^\star/\alpha_\text{N}$. Both models are trained using the dataset introduced in Section\,\ref{Case3}, where only the relevant two properties of each data point are considered. The good performance of these two inverse models are demonstrated by the good agreement between the designed and target properties in Figs. S.8 and S.9.

Inverse model 5 is employed to solve representative inverse design tasks, as listed in Table\,\ref{inverse5}. The corresponding results for inverse model 6 are found in Table\,S.3.
In the first example, the target properties are directly specified, i.e., $\nu^{\star}_{xy}=-1.0$ and $\alpha_{x}^\star/\alpha_\text{N}=-1.0$. In the other examples, the inverse model is used to identify the minimum and maximum values of $\alpha_{x}^\star/\alpha_\text{N}$ for $\nu^{\star}_{xy}=-1.0$, and those of $\nu^{\star}_{xy}$ for $\alpha_{x}^\star/\alpha_\text{N}=-1.0$. These tasks are performed using the search method introduced in Case 2 in Section\,\ref{Case2}. In all examples, inverse model 5 successfully designs the corresponding structures with a relative error in properties smaller than 5\%. These two inverse models are suitable for inverse design tasks involve only partially specified target properties and fixed base material-related features. 
\begin{table}[htbp]
	\caption{Inverse design results obtained using inverse ML model 5.  The first design example corresponds to the case where $\nu^{\star}_{xy}$ and $\alpha_{x}^\star/\alpha_\text{N}$ are specified. In the other four examples demonstrate the capability of the inverse model to determine the minimum and maximum values of one of the two properties when the other is specified.}
	{\small
		\begin{tabular}{p{0.25\textwidth}p{0.25\textwidth}p{0.3\textwidth}p{0.2\textwidth}}
			%\begin{tabular}{lllll}
			\toprule
			Target properties: \newline$\nu^{\star}_{xy}$,\,$\alpha_{x}^\star/\alpha_\text{N}$ & Designed properties:\newline$\nu^{\star}_{xy}$,\,$\alpha_{x}^\star/\alpha_\text{N}$&Designed features: \newline  $\rho$,\, $h/w$,\, $\theta_1$, \, $\theta_2$ & Relative error\\
			\midrule
			-1.0,\,-1.0 &-1.020,\,-1.046 &0.243,\,1.093,\newline$64.7^\circ$,\,$52.9^\circ$
			&3.6\%\\ 
			-1.0,\, minimum &-0.962,\,-1.471 &0.222,\,1.092,\newline$60.4^\circ$,\,$52.8^\circ$
			&3.8\%\newline(w.r.t. $\nu^{\star}_{xy}$)\\	
			-1.0,\, maximum &-1.041,\,-0.408 &0.282,\,1.12,\newline$75.0^\circ$,\,$53.9^\circ$
			&4.1\%\newline(w.r.t. $\nu^{\star}_{xy}$)\\	
			minimum,\, -1.0 &-5.123,\,-0.953 &0.137,\,0.523,\newline$76.0^\circ$,\,$32.6^\circ$
			&4.7\%\newline(w.r.t. $\alpha_{x}^\star/\alpha_\text{N}$)\\	
			maximum,\, -1.0 &-0.184,\,-0.982&0.316,\,1.794,\newline$56.8^\circ$,\,$79.6^\circ$
			&1.8\%\newline(w.r.t. $\alpha_{x}^\star/\alpha_\text{N}$)\\		
			\bottomrule
		\end{tabular}
	}
	\label{inverse5}
\end{table}

In summary, the trained inverse ML models and the proposed methods above are capable of addressing inverse design tasks relevant to the practical application of the hybrid-honeycomb structures. They enable efficient and accurate identification of structures with fully or partially specified target properties.
\section{Conclusion}
\label{conclusion}
This work is devoted to the machine learning-enabled forward prediction and inverse design of a hybrid-honeycomb structure belonging to thermoelastic materials with widely tunable thermoelastic properties. The structure is characterized by four geometrical features and two base material-related features and its effective thermoelastic properties include effective Young's moduli, shear moduli, Poisson's ratios and thermal expansion coefficients. 

A large dataset consisting of randomly generated structures paired with the corresponding effective properties is generated using the computational homogenization method. First, a forward ML model is trained, which efficiently and accurately predicts the effective properties of a structure with given structural features. Subsequently, six inverse ML models corresponding to different practical scenarios where different combinations of effective properties and structural features are considered as design targets and design variables. To address nonuniqueness issue that multiple structures may exhibit the same effective properties, the inverse ML models are trained by minimizing the reconstruction loss, which quantifies the difference between the target properties and those of the designed structures and is evaluated by using the trained forward ML model. The inverse ML models 4, 5, and 6 applicable for cases of with fixed base materials are trained by a new dataset generated by the forward ML model. In all six cases, the inverse model can efficiently predict a structure with properties closely matching the specified target values. Furthermore, they can be applied to conditional optimization tasks when combined with a search method, such as identifying the minimum or maximum value of one property when another is specified. 

The developed forward and inverse ML models, exhibiting excellent predictive and design performance, facilitate the practical application of the hybrid-honeycomb structure as functional/structural components in advanced engineering systems. The proposed ML-based framework can be extended to realize the forward and inverse design of other types of thermoelastic metamaterials.
\section*{Availability of data}
The data and codes required to reproduce findings of this work are found at \url{https://github.com/XiangLongPeng/ML_inverse_design_thermo_metamater.git}.

\section*{Acknowledgments}
The computing time granted on the Hessian High-Performance Computer ‘‘Lichtenberg” and the support from German Research Foundation (DFG) under Grant No. 562095270 are gratefully acknowledged.
%%\bibliography{mybibfile}
\bibliographystyle{model3-num-names}

\end{document}

\begin{frontmatter}

\title{Supplementary Material: \\Machine learning-enabled inverse design of bimaterial thermoelastic lattice metamaterials}
\author[tud]{Xiang-Long Peng\corref{cor1}}
\ead{xianglong.peng@tu-darmstadt.de}
\author[tud]{Bai-Xiang Xu}
\cortext[cor1]{Corresponding author}
\address[tud]{Mechanics of Functional Materials Division, Institute of Materials Science, Technische Universität Darmstadt, 64287 Darmstadt, Germany}

\end{frontmatter}
\renewcommand{\thefigure}{S.\arabic{figure}}
\renewcommand{\thetable}{S.\arabic{table}}
\renewcommand{\theequation}{S.\arabic{equation}}
\section{Hyperparameters for forward and inverse ML models}
The final hyperparameters used for the forward ML model and the six inverse models are listed in Table\,\ref{NN_p2}. These hyperparameters were determined through iterative testing and enable the trained ML models to achieve satisfactory performance.
\begin{sidewaystable}[htbp]
	\caption{Summary of hyperparameters, model architectures, and training settings for the forward and inverse ML models. }
	{\small
		\begin{tabular}{lllllll}
			%{p{0.3\textwidth}p{0.3\textwidth}p{0.4\textwidth}}
			%\begin{tabular}{lll}
			\toprule
			ML  Model & Neural network layers  & Activation function &Optimization algorithm&Learning rate&Epoch number&Batch size\\
			\midrule
			Forward& 6,\,128,\,128,\,128,\,64,\,32,\,6  &leaky relu &Adam& 0.001&1500&500\\
			Inverse 1& 6,\,128,\,128,\,128,\,128,\,6  &leaky relu, \,sigmoid (output layer) &Adam& 0.001&1500&500\\
			Inverse 2& 2,\,128,\,128,\,128,\,6  &leaky relu,\, sigmoid (output layer) &Adam& 0.001&200&500\\		
			Inverse 3& 2,\,128,\,128,\,128,\,6  &leaky relu,\, sigmoid (output layer) &Adam& 0.001&200&500\\		
			Inverse 4& 6,\,128,\,128,\,128,\,4  &leaky relu,\, sigmoid (output layer) &Adam& 0.001&500&500\\			
			Inverse 5& 2,\,128,\,128,\,128,\,4  &leaky relu,\, sigmoid (output layer) &Adam& 0.001&200&500\\		
			Inverse 6& 2,\,128,\,128,\,128,\,4  &leaky relu,\, sigmoid (output layer) &Adam& 0.001&200&500\\										
			\bottomrule
		\end{tabular}
	}
	\label{NN_p2}
\end{sidewaystable}
\section{Loss-epoch curves}
\begin{figure}[!htbp]
	\centerline{\includegraphics[width=0.7\textwidth]{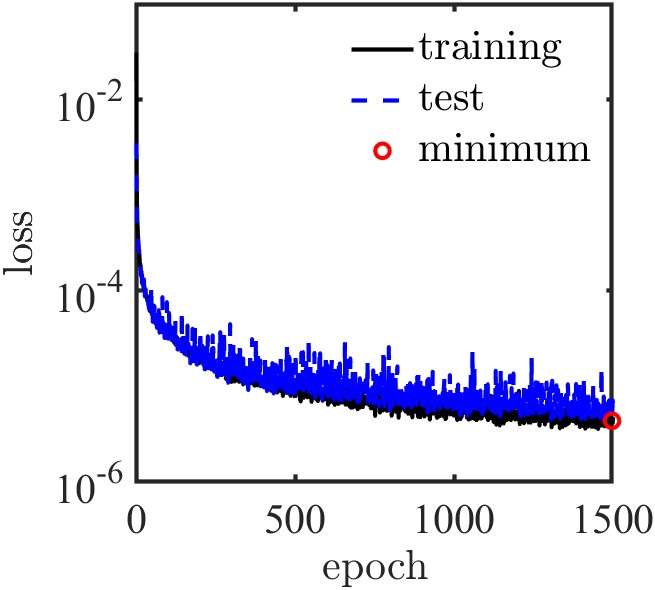}}
	\caption{Loss versus epoch curves for the forward ML model. The model at the epoch corresponding to the smallest test loss (marked by the red circle) is selected as the final, converged model.}
	\label{loss_epoch_forward}
\end{figure}

\begin{figure}[!htbp]
	\centerline{\includegraphics[width=\textwidth]{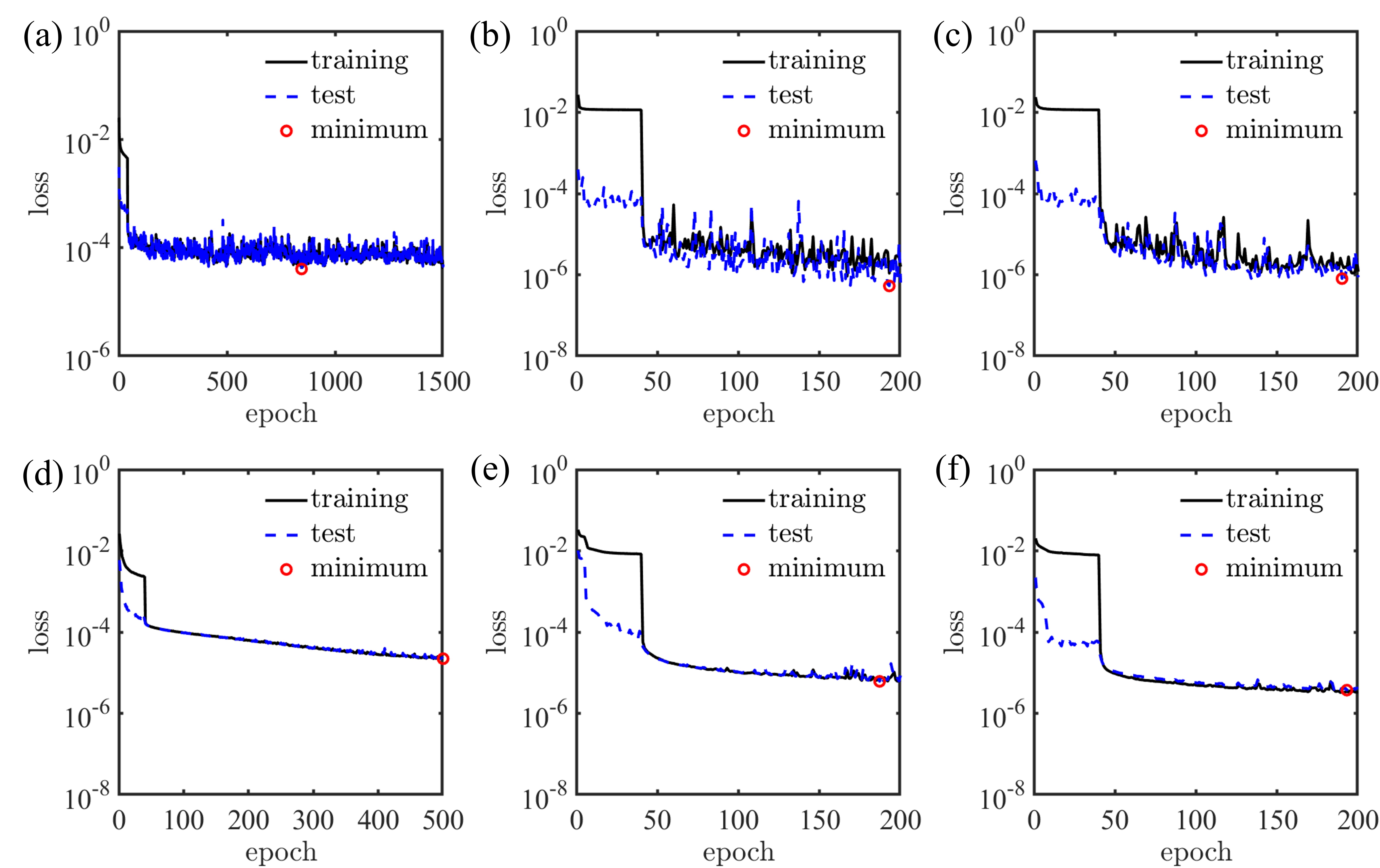}}
	\caption{Loss versus epoch curves for inverse models: (a) to (f) for inverse models 1 to 6. In each case, the model at the epoch corresponding to the smallest test loss (marked by the red circle) is selected as the final, converged model.}
	\label{loss_epoch_inverse}
\end{figure}

The training and test loss versus epoch curves for the forward model are shown in Fig.\,\ref{loss_epoch_forward}, while those for the six inverse models are presented in Figs.\,\ref{loss_epoch_inverse}a-f. The close agreement between the training and test losses indicates that the ML models are properly trained, without significant overfitting or underfitting. The oscillations observed in the loss curves suggest that the final epoch does not necessarily correspond to the minimum loss. Therefore, the ML models are saved after each epoch, and the one achieving the smallest test loss is selected as the final, converged model.
\section{Dataset for inverse ML model training in cases with specified material-related features}
\begin{figure}[!htbp]
	\centerline{\includegraphics[width=\textwidth]{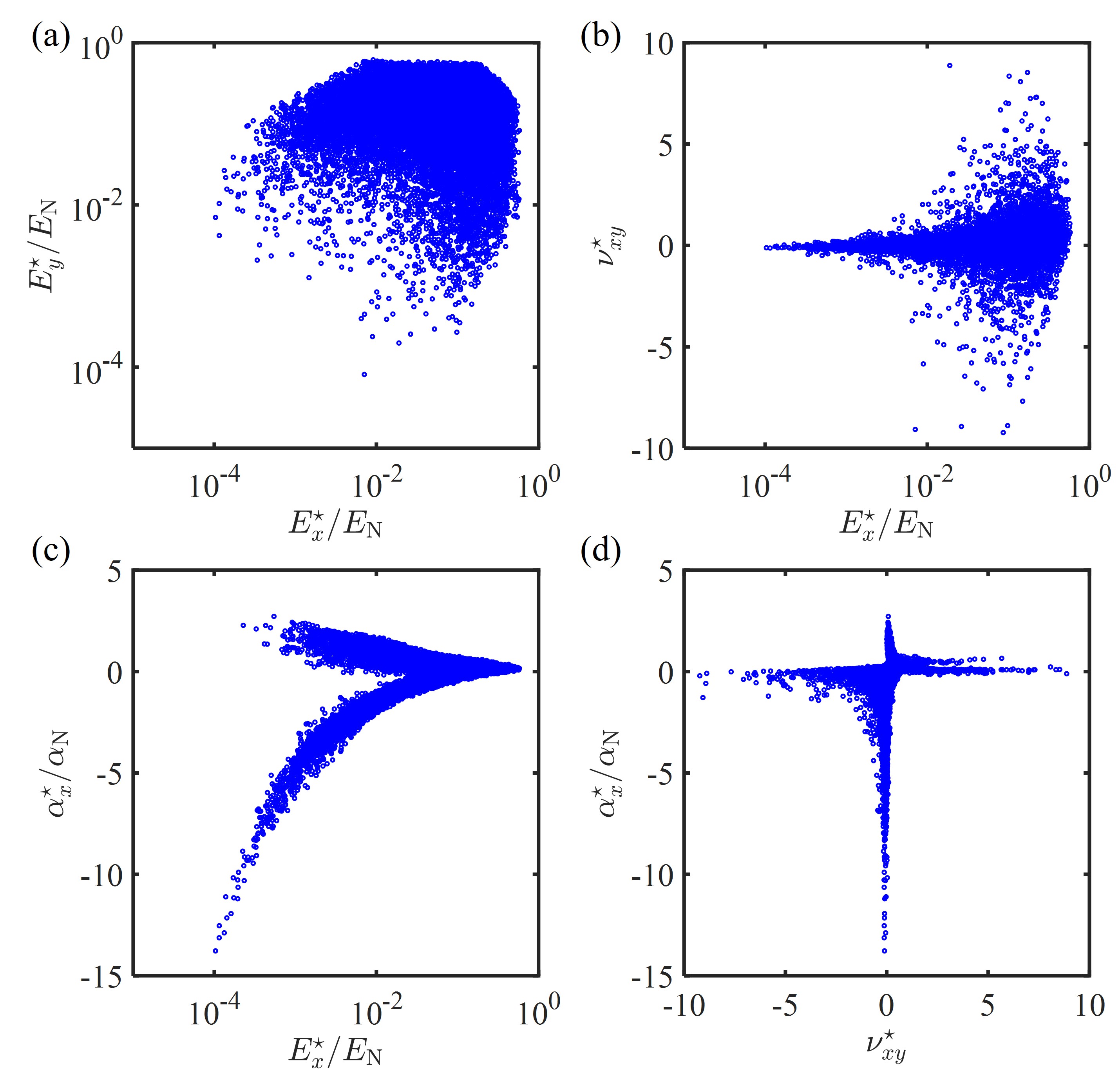}}
	\caption{Effective thermoelastic properties of structures in the dataset with fixed material-related features:  (a) $E^{\star}_{x}/E_\text{N}$ versus $E^{\star}_{y}/E_\text{N}$, (b) $E^{\star}_{x}/E_\text{N}$ versus $\nu^{\star}_{xy}$, (c) $E^{\star}_{x}/E_\text{N}$ versus $\alpha^{\star}_{x}/\alpha_\text{N}$, and (d) $\nu^{\star}_{xy}$ versus $\alpha^{\star}_{x}/\alpha_\text{N}$. }
	\label{partial_dataset}
\end{figure}

The dataset generated by the forward ML model, with the two material-related features being fixed (i.e., $E_\text{I}/E_\text{N}=1.977$ and $\alpha_\text{I}/\alpha_\text{N}=0.0652$), is visualized in Fig.\,\ref{partial_dataset}. This dataset is used to train and test inverse ML models 4, 5, and 6.
\section{Supplementary results for performance of inverse ML models}
\begin{figure}[!htbp]
	\centerline{\includegraphics[width=\textwidth]{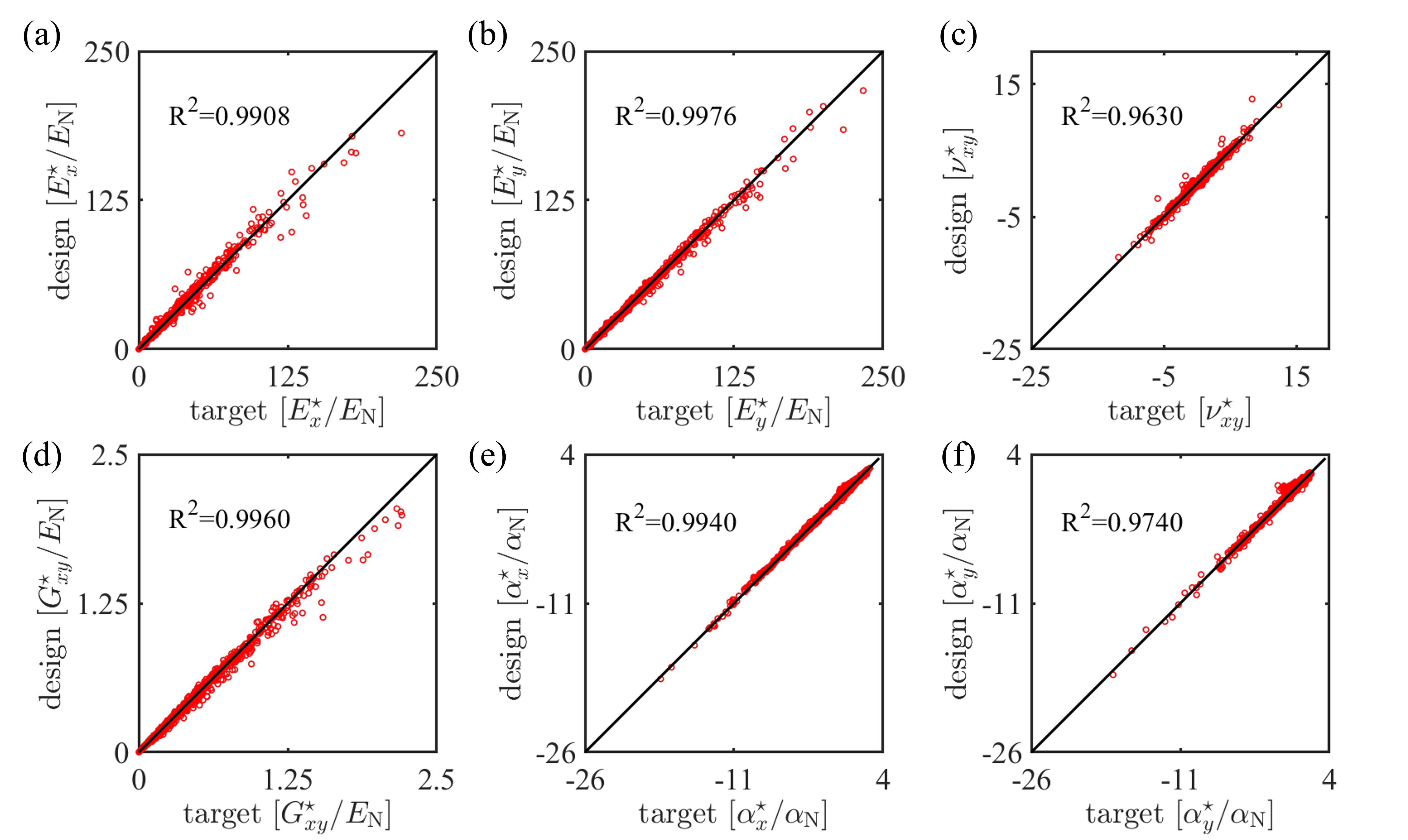}}
	\caption{Properties of structures inversely designed by inverse model 1 versus the target properties in the test dataset: (a) $E^{\star}_{x}/E_\text{N}$, (b) $E^{\star}_{y}/E_\text{N}$, (c) $\nu^{\star}_{xy}$, (d) $G^{\star}_{xy}/E_\text{N}$, (e) $\alpha^{\star}_{x}/\alpha_\text{N}$, and (f) $\alpha^{\star}_{y}/\alpha_\text{N}$. }
	\label{inverse1_R_square}
\end{figure}

\begin{figure}[!htbp]
	\centerline{\includegraphics[width=\textwidth]{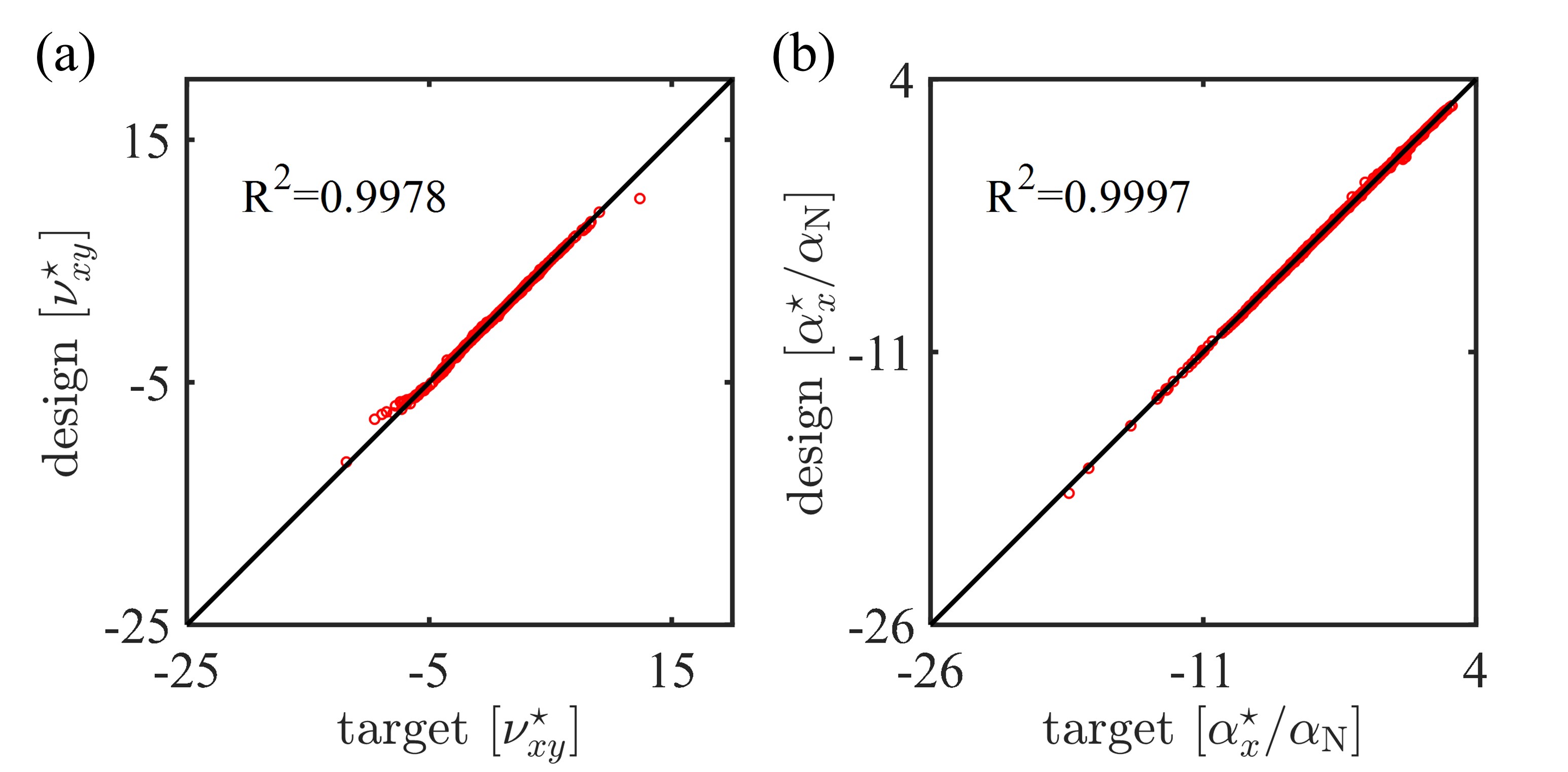}}
	\caption{Properties of structures inversely designed by inverse model 2 versus the target properties in the test dataset: (a) $\nu^{\star}_{xy}$ and (b) $\alpha^{\star}_{x}/\alpha_\text{N}$.}
	\label{inverse2_R_square}
\end{figure}

\begin{figure}[!htbp]
	\centerline{\includegraphics[width=\textwidth]{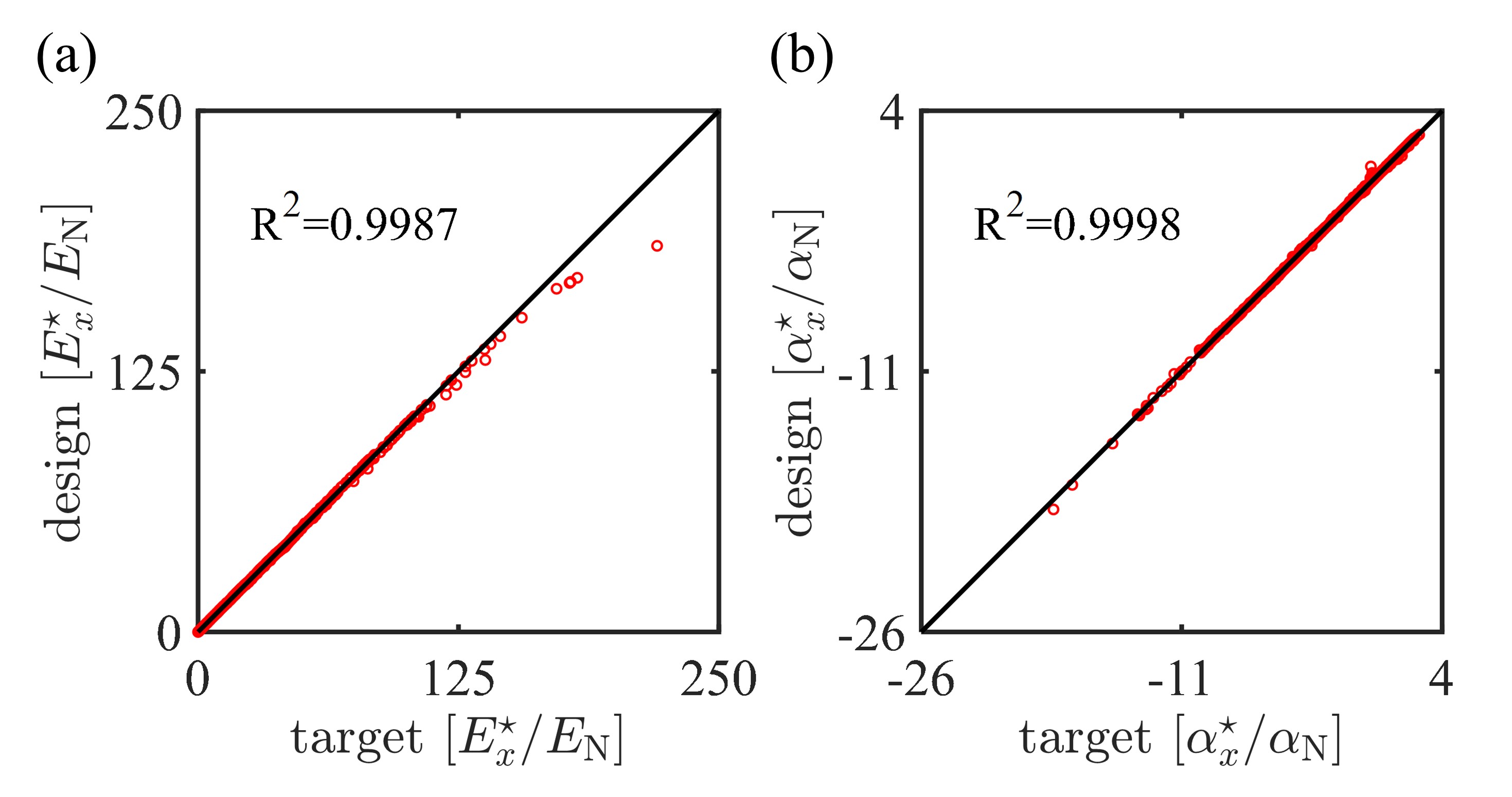}}
	\caption{Properties of structures inversely designed by inverse model 3 versus the target properties in the test dataset: (a) $E^{\star}_{x}/E_\text{N}$ and (b) $\alpha^{\star}_{x}/\alpha_\text{N}$. }
	\label{inverse3_R_square}
\end{figure}

\begin{figure}[!htbp]
	\centerline{\includegraphics[width=\textwidth]{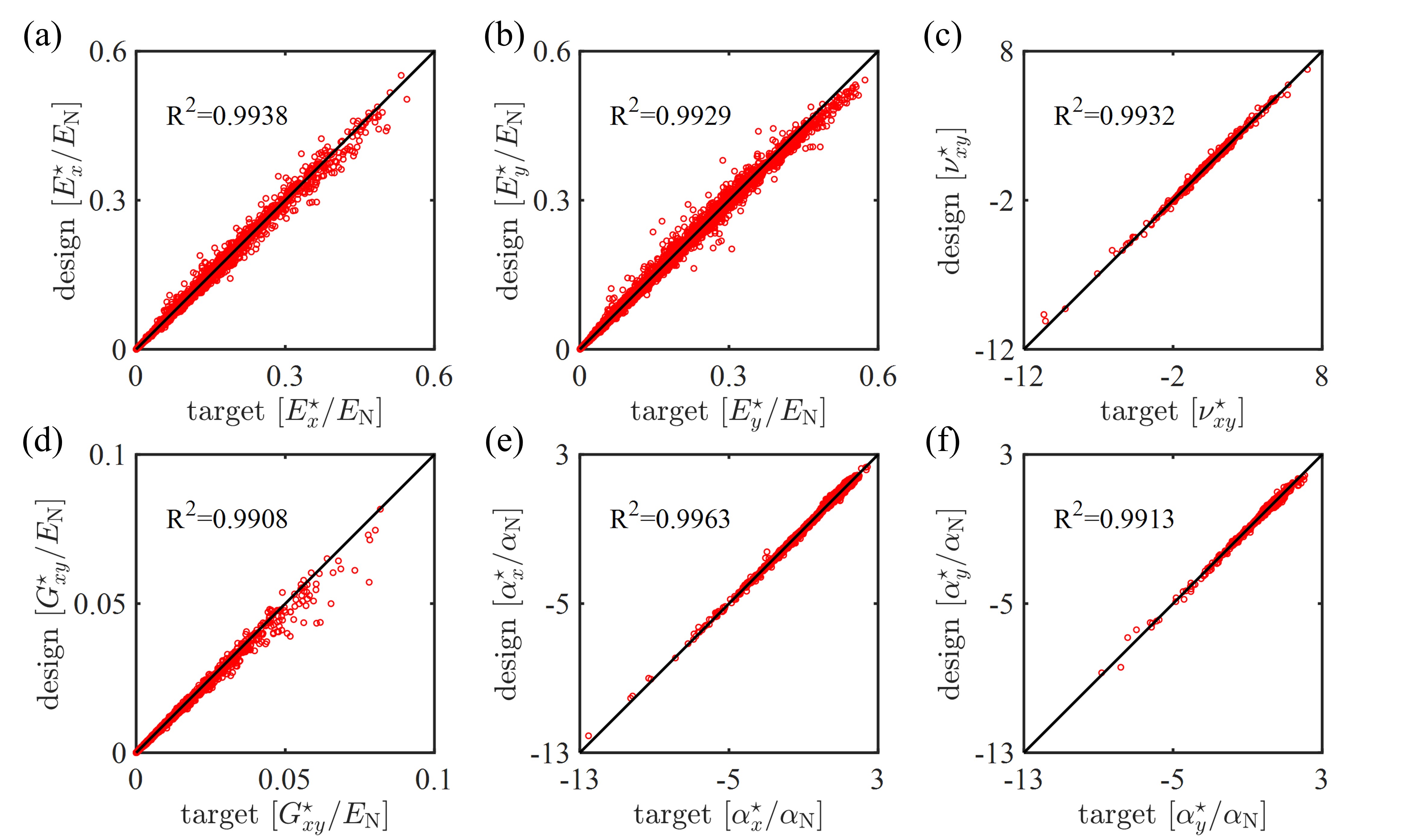}}
	\caption{Properties of structures inversely designed by inverse model 4 versus the target properties in the test dataset: (a) $E^{\star}_{x}/E_\text{N}$, (b) $E^{\star}_{y}/E_\text{N}$, (c) $\nu^{\star}_{xy}$, (d) $G^{\star}_{xy}/E_\text{N}$, (e) $\alpha^{\star}_{x}/\alpha_\text{N}$, and (f) $\alpha^{\star}_{y}/\alpha_\text{N}$. In this case, the design variables are the four geometrical features.}
	\label{inverse4_R_square}
\end{figure}

\begin{figure}[!htbp]
	\centerline{\includegraphics[width=\textwidth]{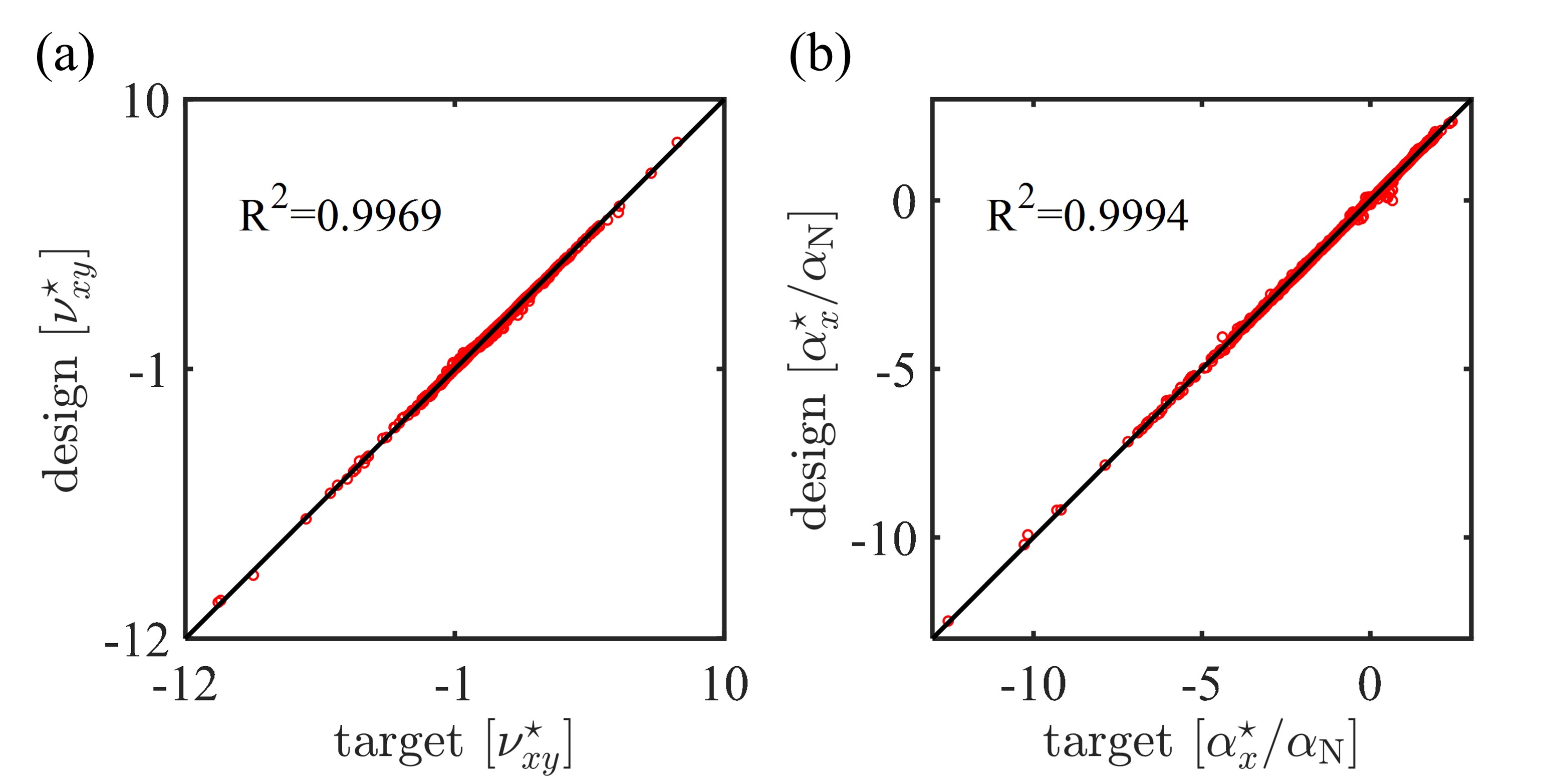}}
	\caption{Properties of structures inversely designed by inverse model 5 versus the target properties in the test dataset: (a) $\nu^{\star}_{xy}$ and (b) $\alpha^{\star}_{x}/\alpha_\text{N}$. In this case, the design variables are the four geometrical features. }
	\label{inverse5_R_square}
\end{figure}

\begin{figure}[!htbp]
	\centerline{\includegraphics[width=\textwidth]{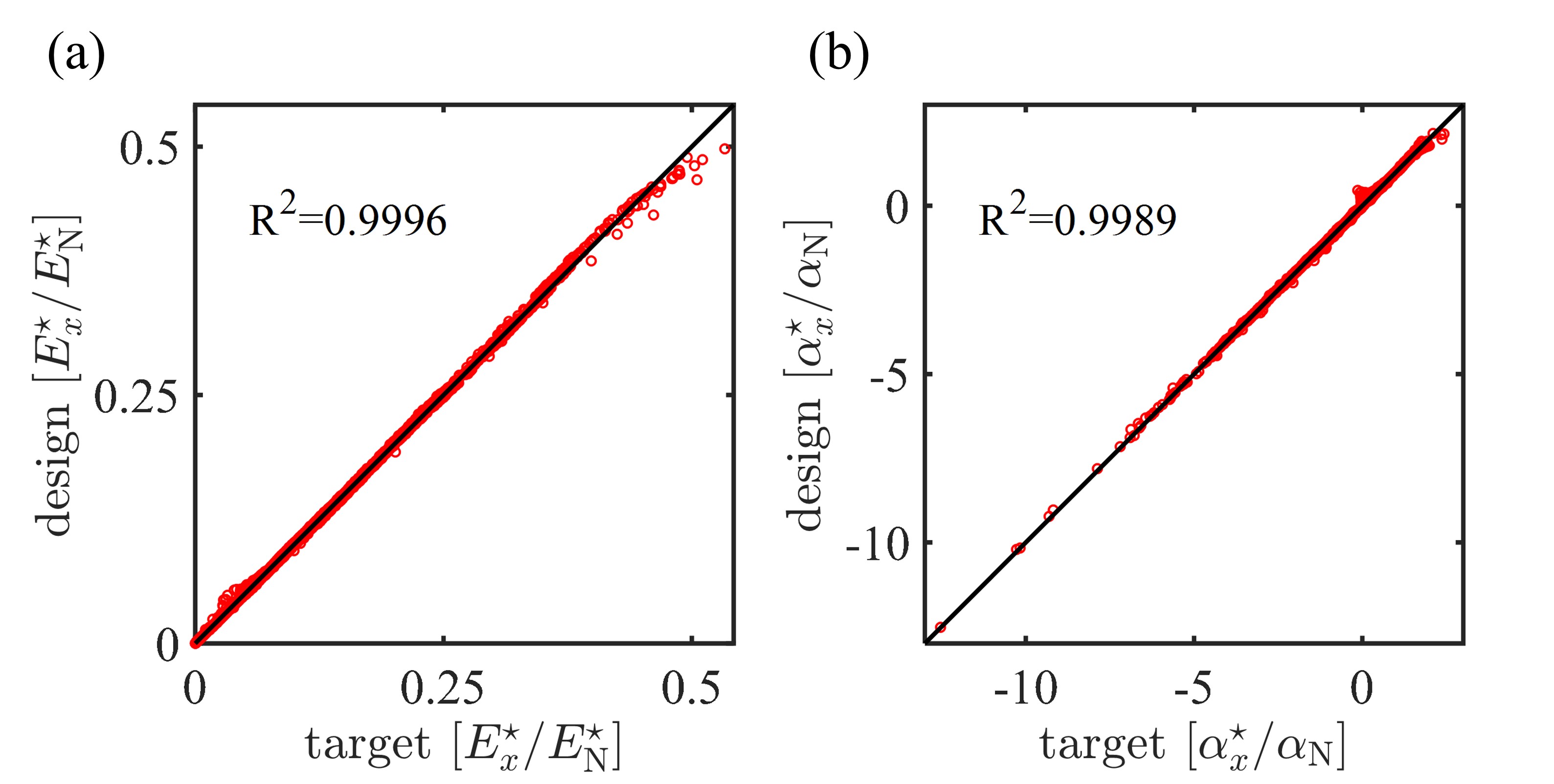}}
	\caption{Properties of structures inversely designed by inverse model 6 versus the target properties in the test dataset: (a) $E^{\star}_{x}/E_\text{N}$ and (b) $\alpha^{\star}_{x}/\alpha_\text{N}$. In this case, the design variables are the four geometrical features.}
	\label{inverse6_R_square}
\end{figure}

The performances of the six inverse ML models on the test dataset are shown in Figs.\,\ref{inverse1_R_square}–\ref{inverse6_R_square}, respectively. The close agreement between the properties of the inversely designed structures and the target properties, as indicated by the high coefficients of determination (i.e., $R^2$ factors), demonstrates the excellent performances of the inverse models.

Representative inverse design examples by inverse models 3 and 6 are presented in Tables\,\ref{inverse3} and \ref{inverse6}, respectively. These results further confirm their good performance.

\begin{table}[htbp]
	\caption{Inverse design results obtained using inverse ML model 3.  }
	{\small
		\begin{tabular}{p{0.25\textwidth}p{0.25\textwidth}p{0.3\textwidth}p{0.2\textwidth}}
			\toprule
			Target properties: \newline$E^{\star}_{x}/E_\text{N}$,\,$\alpha_{x}^\star/\alpha_\text{N}$ & Designed properties:\newline$E^{\star}_{x}/E_\text{N}$,\,$\alpha_{x}^\star/\alpha_\text{N}$&Designed features: \newline $E_\text{I}/E_\text{N}$,\, $\alpha_\text{I}/\alpha_\text{N}$,\newline $\rho$,\, $h/w$,\, $\theta_1$, \, $\theta_2$ & Relative error\\
			\midrule
			0.01,\,-1.0 &0.01,\,-0.987 &0.333,\,0.03,\newline0.294,\,1.82,\newline$60.1^\circ$,\,$97.7^\circ$
			&1.3\%\\ 
			0.01,\, minimum &0.00984,\,-2.023 &1.46,\,0.00458,\newline0.348,\,2.15,\newline$41.5^\circ$,\,$101.3^\circ$
			&1.6\%\newline(w.r.t. $E^{\star}_{x}/E_\text{N}$)\\	
			0.01,\, maximum &0.01002,\,1.41 &1.78,\,0.0459,\newline0.302,\,2.75,\newline$135.6^\circ$,\,$87.6^\circ$
			&0.2\%\newline(w.r.t. $E^{\star}_{x}/E_\text{N}$)\\	
			\bottomrule
		\end{tabular}
	}
	\label{inverse3}
\end{table}

\begin{table}[htbp]
	\caption{Inverse design results obtained using inverse ML model 6.  }
	{\small
		\begin{tabular}{p{0.25\textwidth}p{0.25\textwidth}p{0.3\textwidth}p{0.2\textwidth}}
			%\begin{tabular}{lllll}
			\toprule
			Target properties: \newline$E^{\star}_{x}/E_\text{N}$,\,$\alpha_{x}^\star/\alpha_\text{N}$ & Designed properties:\newline$E^{\star}_{x}/E_\text{N}$,\,$\alpha_{x}^\star/\alpha_\text{N}$&Designed features: \newline $\rho$,\, $h/w$,\, $\theta_1$, \, $\theta_2$ & Relative error\\
			\midrule
			0.015,\,-1.5 &0.0153,\,-1.502 &0.369,\,2.474,\newline$45.0^\circ$,\,$104.6^\circ$
			&0.15\%\\ 
			0.015,\, minimum &0.0146,\,-1.646 &0.430,\,2.43,\newline$40.1^\circ$,\,$106.2^\circ$
			&2.7\%\newline(w.r.t. $E^{\star}_{x}/E_\text{N}$)\\	
			0.015,\, maximum &0.0153,\,1.390 &0.318,\,2.59,\newline$135.7^\circ$,\,$74.6^\circ$
			&2\%\newline(w.r.t. $E^{\star}_{x}/E_\text{N}$)\\		
			\bottomrule
		\end{tabular}
	}
	\label{inverse6}
\end{table}